\documentstyle[preprint,aps]{revtex}
\preprint{IFUSP-P/1137}
\begin{document}
\draft
\title{Comment on ``A New Symmetry for QED'' and \\
``Relativistically Covariant Symmetry in QED'' }
\author{Victor O. Rivelles \cite{cnpq}}
\address{Instituto de F\'\i sica, Universidade de S\~ao Paulo, \\
C.Postal 20516, CEP 01498-970, S\~ao Paulo, SP, Brazil}
\maketitle

\begin{abstract}
We show that recently found symmetries in QED are just non-local
versions of standard BRST symmetry.
\end{abstract}

\pacs{11.10.Ef, 12.20.-m}

Recently it was found a non-local and non-covariant symmetry of QED
in the Feynman gauge
by Lavelle and McMullan \cite{lm} which was cast in a covariant form
by Tang and Finkelstein \cite{tf}. It was claimed that they are new
symmetries of QED and give rise to new Ward identities. We would like
to point out that these symmetries are standard BRST symmetries and
therefore they can not give rise to any new Ward identity.

In the Hamiltonian formulation of QED besides the gauge field and its
momenta we have a pair of ghost fields (${c, \overline{c}}$) and its
momenta (${\cal \overline{P}, P}$) \cite{henneaux} (we leave out the
fermion fields since they are not essential for our purposes and can be
easily included). The ghost Lagrangian of QED (still in the Hamiltonian
form) which implements the
Lorentz condition is then found to be \cite{henneaux} $L^{(ham)}_{gh} =
\dot{\cal P} \overline{c} + \dot{c} \overline{\cal P} -
i \overline{c} \nabla^2 c + i \overline{\cal P} {\cal P}$. Usually
the next step is to perform the integration over the ghost momenta to
get the usual ghost Lagrangian $L_{gh} = i \overline{c} \Box c$.
However we can perform the integration over the ghost fields instead
of their momenta. Performing the integration over $\overline{c}$ we
get a delta functional $\delta( i \nabla^2 c + \dot{\cal P} ) =
det \nabla^2 \,\,\, \delta( i c + \frac{1}{\nabla^2} \dot{\cal P} )$. Now
performing the integration over $c$ we get the non-local ghost
Lagrangian $L^{(non-local)}_{gh} = - i \overline{\cal P}
\frac{1}{\nabla^2} \ddot{\cal P} +
i \overline{\cal P} {\cal P}$ and the non-local BRST transformations
$ \delta A_i = i \frac{\partial_i}{\nabla^2} \dot{\cal P}, \,\,
\delta A_0 = i {\cal P}, \,\, \delta {\cal P} = 0, \,\,
\delta \overline{\cal P} = \nabla^2 A_0 - \partial_i \dot A_i $. We can now
perform the following change of variables $ \overline{\cal P} =
\nabla^2 \overline{\cal R}$ in order to get a local action and to
get rid of the term $ det \nabla^2 $ in the path integral measure
(which came from the integration over $\overline{c}$). After this
change of variables we get the usual ghost action $L_{gh} =
i \overline c \Box c$ and the non-covariant and non-local
transformations of Lavelle and McMullan \cite{lm} after identifying
$\cal P$ with $\overline c$ and $\overline{\cal R}$ with $c$.
Since we have a standard BRST symmetry we get the usual constraints
on the physical states and no further independent Ward identities
can be found.

We now turn to the Tang and Finkelstein transformations. The ghost
Lagrangian of QED $L_{gh} = i \overline{c} \Box c$
has a huge freedom when we perform field
redefinitions in the ghost fields $c$ and $\overline{c}$.
%The only restriction is that these
%field redefinitions leave the path integral measure invariant.
If we consider, e.g., the following non-local redefinitions
$
c = \frac{1}{\nabla^2}\partial_0 d,  \,\,
\overline{c} = \frac{1}{\partial_0} \nabla^2 \overline{d}
$,
the Lagrangian and the path integral measure remain invariant and
the usual BRST transformations become
$
\delta A_\mu = \partial_\mu \frac{1}{\nabla^2} \partial_0 d,\,\,
\delta d = 0, \,\, \delta \overline{d} = - \frac{i}{\xi}
\frac{1}{\nabla^2} \partial_0 \partial_\mu A^\mu
%\delta \psi = g \psi \frac{1}{\nabla^2} \dot d  ,
%\delta \overline{\psi} = g \frac{1}{\nabla^2} \dot d \overline{\psi}
$.
These are the covariant non-local transformations presented in Ref.\cite{tf}
(written in an arbitrary gauge, i.e., arbitrary $\xi$)
after identifying $d$ and $\overline{d}$ with $\overline{c}$ and $c$
respectively. Of course
this procedure can be generalized to any (local or non-local)
redefinition of the ghost fields which leave the action and the path
integral measure invariant.

The gauge fixed QED action is also invariant under anti-BRST transformations
which anticommute with the BRST transformations. We can then perform
an arbitrary field redefinition (which leaves the action and the
path integral measure invariant) and consider the BRST
transformations of the redefined fields. Then perform a second arbitrary
field redefinition and consider  the anti-BRST transformations of the
redefined fields. Since the original action is invariant under these
field redefinitions the BRST and anti-BRST transformations of the
redefined fields are still symmetries of the action.
The sum of these two transformations are precisely the Tang and
Finkelstein transformations Eqs.(5) and (9).
Originally the BRST and anti-BRST transformations are anticommutating but
now, since they are acting after field redefinitions they no longer
need to anticommute. This explains why the transformations Eqs.(5) and
(9) of Ref.\cite{tf} are no longer nilpotent. In fact the anticommutator
gives rise to a new field redefinition which is also a symmetry of
the action as can be easily  verified.


\begin{references}
\bibitem[*]{cnpq}Partially supported by CNPq. E-mail: vrivelles@if.usp.br.
\bibitem{lm}M. Lavelle and D. McMullan, Phys. Rev. Lett. {\bf 71}, 3758
(1993).
\bibitem{tf}Z. Tang and D. Finkelstein, Phys. Rev. Lett. {\bf 73}, 3055
(1994).
\bibitem{henneaux} M. Henneaux, Phys. Rep. {\bf 126}, 1 (1985).
\end{references}
\end{document}